# Temperature Anisotropies in a Universe with Global Defects *

David Coulson
*The Blackett Laboratory, Imperial College, London SW7 2BZ, UK,*
and
*Joseph Henry Laboratory, Princeton University,*
Princeton NJ, 08544.

July 15, 1994

## Abstract

We present a technique of calculating microwave anisotropies from global defects in a reionised universe. We concentrate on angular scales down to one degree where we expect the nongaussianity of the temperature anisotropy in these models to become apparent.

## 1  Introduction

The formation of structure in the universe remains one of the most intriguing open questions of science. Over the past decade a number of very different physical mechanisms have been suggested as candidates to explain the universe we observe today. Certainly the most well-understood of these models is the "Standard" CDM scenario [1]. This hypothesises an adiabatic universe with gaussian fluctuations generated at very early epochs when inflation stretched microscopic quantum fluctuations to very large length scales.

This model is very appealing; a number of cosmological problems are naturally solved by inflation, the power spectrum of density fluctuations has roughly the correct form, and definite predictions can be made about the structure of fluctuations in the microwave background. The predicted

---

gaussian fluctuations on large angular scales have a scale invariant spectrum, while on smaller scales (of order the horizon at last scattering) the adiabatic oscillations of the photon-baryon fluid leave a complicated signature in the spectrum [2].

COBE's detection of gaussian fluctuations in the microwave background with an approximately scale invariant power spectrum on angular scales $\gtrsim 7^o$ [3] was hailed as a great success of the inflationary scenario. Recent work, however, has showed that many of the features seen in the COBE maps may equally well result from another set of theories of structure formation, based on a simple but wholly different physical model, that of cosmic defects.

According to these theories, symmetries are broken as the hot early Universe cooled, forming a disordered phase. This phase would contain defects such as cosmic strings, monopoles or textures [4]. As the universe expands, the defects begin to order themselves on progressively larger scales, inducing *time-dependent* perturbations in the matter and metric fields of the early universe.

The evolution of these defect fields is highly non-linear, producing a nongaussian pattern of fluctuations. As microwave photons pass through these evolving defect fields, they receive energy shifts, leaving a nongaussian signature on the microwave sky. However, on large angular scales (such as those probed by COBE), the temperature anisotropy pattern is predicted to be quite gaussian, and scale invariant [5], [6], [7]. In these *Proceedings* we describe a technique for determining typical temperature anisotropy maps on angular scales down to $\simeq 1^o$ for global defect models with reionisation. Using these maps we present a test for nongaussianity which may ultimately distinguish defect-induced patterns from gaussian noise.

Throughout these *Proceedings* we will assume the canonical values for cosmological parameters: $\Omega = 1$, $\Lambda = 0$, $h = 0.5$ and $\Omega_{baryon} = 0.05$.

## 2  Sources of CMBR Anisotropy

Depending on the thermal history of the universe, the baryons and electrons of the early universe recombine and become neutral at some epoch $Z_{rec}$, when the photons Thomson scatter for the last time off free electrons. In a universe with no reionisation, the photons then travel freely to the observer, red-shifting from the expansion of the universe. Any observed anisotropy in the temperature of these photons gives us information about both the physics at the epoch of recombination, and the environment through which the photons passed from the surface of last scattering to us.

The final temperature anisotropy for a perturbed flat FRW universe, with metric given by $g_{\mu\nu} = a^2(\eta)(\eta_{\mu\nu} + h_{\mu\nu})$, can be written as

$$\left[\frac{\Delta T}{T}\right]_f = \left[\frac{\Delta T}{T}\right]_i - \frac{1}{2}\int_i^f h_{\mu\nu,0}(x^\alpha_{(0)}(\eta))\hat{n}^\mu\hat{n}^\nu \, d\eta + \left[\frac{1}{2}h_{00}\right]_i^f + [\delta\mathbf{v}\cdot\hat{\mathbf{n}}]_i, \quad (1)$$

where $x^\alpha_{(0)} = n^\alpha\eta$ is the unperturbed photon geodesic, with $n^\alpha = (1, -\mathbf{n})$ and $\mathbf{n}^2 = 1$, and $\delta\mathbf{v}$ is the peculiar velocity of the photon fluid [6]. We have removed the dipole contribution to the anisotropy from our own peculiar motion.

The expression $[\Delta T/T]_i$ is the intrinsic temperature anisotropy on the surface of last scattering, and is given by

$$\left[\frac{\Delta T}{T}\right]_i = \frac{1}{4}\delta_r, \quad (2)$$

where $\delta_r$ is the perturbation in the radiation density.

Equation 1 relates the temperature anisotropy, $\Delta T$, in some direction $\hat{\mathbf{n}}$ on the sky, at conformal time $\eta_f$, with the anisotropy at an initial epoch, $\eta_i$, the epoch of last scattering. Working in synchronous gauge, where the metric perturbations are constrained by $h_{00} = h_{0i} = 0$, the anisotropy is simply given by

$$\left[\frac{\Delta T}{T}\right]_f = \left[\frac{\Delta T}{T}\right]_i - \frac{1}{2}\int_i^f h_{ij,0}\hat{n}^i\hat{n}^j \, d\eta + [\delta\mathbf{v}\cdot\hat{\mathbf{n}}]_i. \quad (3)$$

According to the standard theory, the epoch of recombination occurs at a redshift $Z_{rec} \simeq 1,100$. However in cosmic defect theories, a substantial fraction of the matter in the universe could have undergone gravitational collapse and formed stars at very high redshifts [8]. These non-linear structures can re-heat the IGM, making reionisation at redshifts higher than 100 possible [6]. The main effect of reionisation on the microwave anisotropy is to mask structure induced at early times, and so to smooth the anisotropy on small angular scales. The characteristic smoothing scale will correspond approximately to the mean free path of the photons at the last scattering surface. In standard scenarios, this is an angular scale of $\simeq 2°$, however, reionisation brings forward the surface of last scattering, increasing the smoothing scale to $\simeq 6°$.

Taking the epoch of last scattering to be $Z_{ls} \simeq 100$ allows us to make a number of approximations in our calculation. First, matter-radiation equality occurs at $Z_{eq} \simeq 23,700h^2 \gg 100$, allowing us to assume matter domination. Further, the baryons can begin to track the CDM after Compton dragging has become negligible at $Z_C \simeq 160h^{\frac{2}{5}}$. For epochs after $Z_C$ we can assume that the baryon velocity has reached the CDM velocity, which in synchronous gauge is given by $v_{cdm} \equiv 0$. Hence, in this gauge there will be no explicit doppler kick from the photons last scattering off the moving electrons, and the term $[\delta \mathbf{v} \cdot \hat{\mathbf{n}}]_i$ of equation 3 vanishes.

## 3  Modelling a Reionised Universe

Although a number of authors have approximated the effect of a reionised universe with a visibility function in the Sachs-Wolfe integral [6], we used a more realistic model of tracing individual photon paths using a Monte Carlo simulation [10].

During the time interval $[t_i, t_f]$, the probability for a photon to scatter is given by

$$P = e^{-d}, \qquad (4)$$

where

$$d = \int_{t_i}^{t_f} dt \; n_e(t) \sigma_T = n_0 \sigma_T t_0 \left[ \frac{t_0}{t_i} - \frac{t_0}{t_f} \right] \qquad (5)$$

and where $t$ is the cosmic time, $n_e(t) = n_0 \, t_0^2/t^2$ is the electron density, and the subscript 0 refers to the present.

Our technique is to construct the photon paths backwards in time, propagating away from the observer. At each time step, we apply equation 5 to find the probability of a scattering event to occur in that time step. We then draw a number at random from a uniform distribution between 0 and 1 to determine whether a scattering actually occurred. The scattering angles are chosen to be isotropic [10].

Having determined the photon's random walk through the baryon plasma, we evaluate the Sachs-Wolfe integral of equation 3 along the scattered path. We follow the derivation of Pen et. al. [6], although find that careful integration by parts along the scattered path produces an extra term at each scattering site. The Sachs-Wolfe integral can be written

$$\left( \frac{\Delta T}{T} \right) = -\frac{1}{2} \left( \int_i^f d\eta \, h_{ij,0}(\eta, \mathbf{x}(\eta)) n^i n^j + \sum_m [\nabla(\chi(\eta, \mathbf{x}(\eta)) \cdot \mathbf{n}]_{i_m}^{f_m} \right), \qquad (6)$$

where $\mathbf{x}(\eta)$ is the photon path, $\mathbf{n}(\eta)$ the photon path direction, and where we have labelled times on the photon path such that the photon travels in a straight line in the time interval $(i_m, f_m)$. The metric perturbation, $h_{ij,0}$, has contributions from the scalar, vector and tensor modes of the defect stress-energy tensor, as described in reference [6], while the field $\chi$ is given by

$$\chi(\eta, \mathbf{x}) = -\frac{1}{45}\eta \int_0^\eta d\eta' \eta' \mathcal{S}(\eta', \mathbf{x}) + \frac{1}{30}\eta^{-4} \int_0^\eta d\eta' \eta'^6 \mathcal{S}(\eta', \mathbf{x})$$
$$+ \frac{1}{18}\eta^{-2} \int_0^\eta d\eta' \eta'^4 \mathcal{S}(\eta', \mathbf{x}). \qquad (7)$$

Here, $\mathcal{S} = -8\pi G(\Theta + \Theta^s)$, where $\Theta$ and $\Theta^s$ are the trace and scalar contributions to the defect field stress-energy tensor. We have seen that in the synchronous gauge, forcing the baryons to trace the CDM means that the baryon velocity vanishes. However, we have recovered a contribution to the Sachs-Wolfe integral which may be identified as a "Doppler term", $[\nabla \chi \cdot \mathbf{n}]$.

## 4   Numerical Implementation

To evolve the defect fields, we use the non-linear sigma model approximation, constraining the defect fields to remain on the vacuum manifold at all times [11]. Discretising the action

$$S = \int d\eta \, a^2(\eta) \frac{1}{2} \left( -\dot{\vec{\psi}}^2 + (\nabla \vec{\psi})^2 + \lambda(\vec{\psi}^2 - 1) \right), \qquad (8)$$

where $\lambda$ is the Lagrange multiplier, on a cubic space-time grid and varying with respect to $\vec{\psi}$ gives a simple set of defect evolution equations described fully in reference [6].

We begin our simulation by first constructing a set of photon paths. A typical run employs a box of $128^3$ for the field evolution and $64^3$ for the metric perturbations and photon paths. There are typically $64^2$ lines of sight, each with 100 photons. (Using fewer photons leads to a white noise contamination of the final maps on pixel scales from the photon shot noise.) The photons converge on an observer on the face of the lattice, where they subtend a square $30°$ on a side. The whole simulation, including photons, defect fields and perturbation variables $h_{ij,0}$ and $\chi$ is evolved forward from $\eta = 0$ to the present. In each simulation we place observers on each face of the lattice. The resulting six maps are not completely independent, but are useful to get a reasonable measure of the cosmic variance.

A number of tests were performed on the code. To test the evolution of the perturbation variables $h_{ij,0}$ and $\chi$ we verified the code reproduced known analytical solutions for defect seeded perturbations in a Minkowski universe. In such a universe, a spherically symmetric, self-similar texture has a particularly simple stress-energy tensor [12], of the form $\Theta_{ij} = \frac{1}{3}\delta_{ij}\Theta$, where

$$\Theta = 6\phi_0^2 \frac{r^2 - (\eta - \eta_0)^2}{(r^2 + (\eta - \eta_0)^2)^2}. \qquad (9)$$

The co-ordinates here are chosen such that the texture unwinds at the spatial origin at a time $\eta = \eta_0$, and where $\phi_0$ is the defect symmetry breaking scale. By solving the perturbation equations analytically, if we begin integrating the perturbation equations at time $\eta_i$, the metric perturbation is given by

$$\dot{h}_{ij} = -\delta_{ij} 32\pi G \phi_0^2 \left( \frac{\eta - \eta_0}{(\eta - \eta_0)^2 + r^2} - \frac{\eta_i - \eta_0}{(\eta_i - \eta_0)^2 + r^2} \right) \qquad (10)$$

while the "velocity potential", $\chi$, is given by

$$\chi = -8\pi G \phi_0^2 \left( -\frac{2}{5}\eta_0^2 \right) \left( \frac{\eta - \eta_0}{(\eta - \eta_0)^2 + r^2} - \frac{\eta_i - \eta_0}{(\eta_i - \eta_0)^2 + r^2} \right). \qquad (11)$$

These expressions were compared to results of a simulation using a $50^3$ grid, in which the spherically symmetric texture with an initial radius of 10 grid units is evolved. The expansion of the universe was not turned off for this test, although the simulation was run at a very high start time, $\eta_i = 4000\Delta\eta$, to simulate a Minkowski universe. At time $\eta = \eta_0 - 5.1$ we compared the changes in $\dot{h}_{11}$ and $\chi$ with the expected differences from expressions 10 and 11. Results are shown in figures 1 and 2.

We see the correspondence between the predicted analytic results and the simulation is very good in a sphere of radius 15 grid units about the centre of the lattice. (The speed of light in the simulation is unity, so after 5 time units of propagation, we should expect the simulation and analytic result to match in a sphere of $10 + 5 = 15$ grid units.)

Other tests were performed to check the final temperature anisotropies from the code. One such test was to turn off the defect evolution algorithm, and put in "by hand" the stress-energy tensor for an oscillating circular loop of cosmic string in a Minkowski universe.

Following Stebbins [13], we assume the cosmic string loop to have a configuration given by:

$$\mathbf{r}(\sigma, t) = \frac{L}{2\pi} \cos\frac{2\pi t}{L} \left( \cos\frac{2\pi\sigma}{L}\hat{\mathbf{x}} + \sin\frac{2\pi\sigma}{L}\hat{\mathbf{y}} \right), \qquad (12)$$

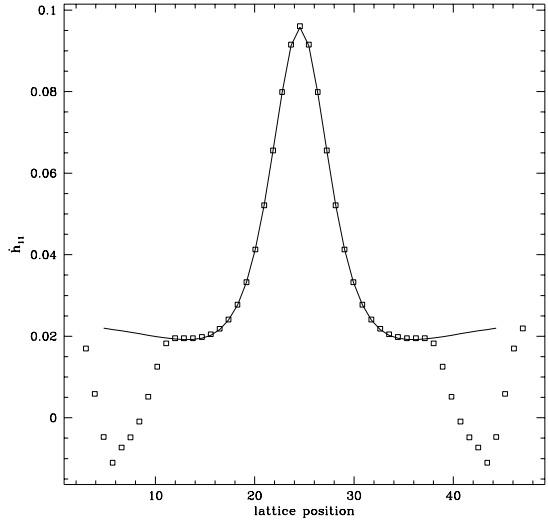

Figure 1: $\dot{h}_{11}$ (in units of $32\pi G\phi_0^2$) for a spherically symmetric texture in a $50^3$ lattice. Boxes are the simulation results, the line is the analytic prediction.

where we have taken the origin of coordinates to be the centre of the circle, the maximum radius of the loop to be $L$, and $t$ and $\sigma$ to parameterise time and the loop position respectively.

The stress-energy of the loop is then

$$\Theta_{\alpha\beta}(\mathbf{x},t) = \tilde{\mu} \oint \begin{pmatrix} 1 & -\dot{\mathbf{r}} \\ -\dot{\mathbf{r}} & -r'_i r'_j \end{pmatrix} \delta^{(3)}(\mathbf{x}-\mathbf{r}(\sigma,t))\, d\sigma \qquad (13)$$

where dots are differentiation with respect to $t$, primes differentiation with respect to $\sigma$, and $\tilde{\mu} \equiv \frac{G\mu}{c^2}$ is proportional to the mass per unit length of the string, $\mu$ [14].

For an observer "face on" to the loop, the predicted temperature anisotropy on the sky is a circular "top hat", given by

$$\frac{\Delta T}{T} = \begin{cases} 8\pi\tilde{\mu}\tan\frac{2\pi\tau}{L} & R_\gamma < |\cos\frac{2\pi\tau}{L}| \\ 0 & R_\gamma > |\cos\frac{2\pi\tau}{L}| \end{cases} \qquad (14)$$

where $R_\gamma$ is the projected radius of the loop, and $\tau$ the impact time of the photons.

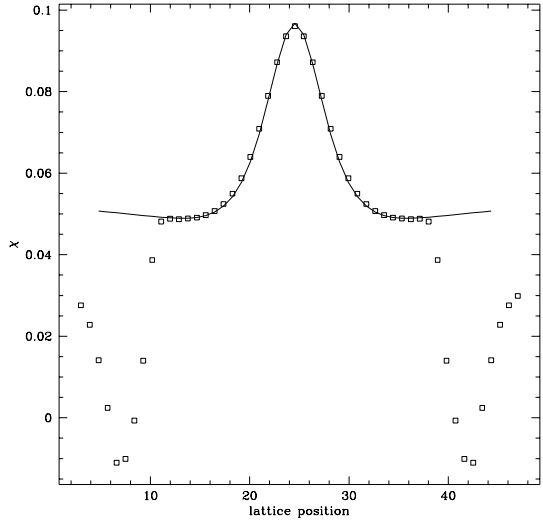

Figure 2: $\chi$ (in units of $\frac{16}{5}\pi G\phi_0^2\eta_0^2$) for a spherically symmetric texture in a $50^3$ lattice. Boxes are the simulation results, the line is the analytic prediction.

Using the analytic stress-energy tensor above, a simulation was run to calculate the temperature anisotropy induced in a plane of photons passing through the loop. By changing the initial phase of the oscillation in the string configuration, (equation 12), we were able to perform several runs, with different impact times. (We used this method rather than starting the plane of photons closer to the loop because equation 14 holds for photon trajectories which begin and end far from the cosmic string. We therefore needed to ensure the photons started and ended as far away from the loop as possible.) A $50^3$ metric perturbation lattice was used, with a $100^3$ defect lattice.

From the final anisotropy maps, the difference between the mean hot or cold spot temperature and the ambient temperature was measured and presented in figure 3. The dashed line is the analytic solution of equation 14. Again, one sees agreement between the predicted analytic expression and the value of $\frac{\Delta T}{T}$ calculated from the simulation. Only when the analytic temperature anisotropy diverges does the simulation not trace it well.

Finally, we were able to verify finite size effects and spatial resolution effects were not a problem by considering the standard deviation of a $(30°)^2$ temperature anisotropy map (smoothed with a Gaussian beam of FWHM 1.5°) using global textures as a source, as a function of the (gravity field) lattice size.

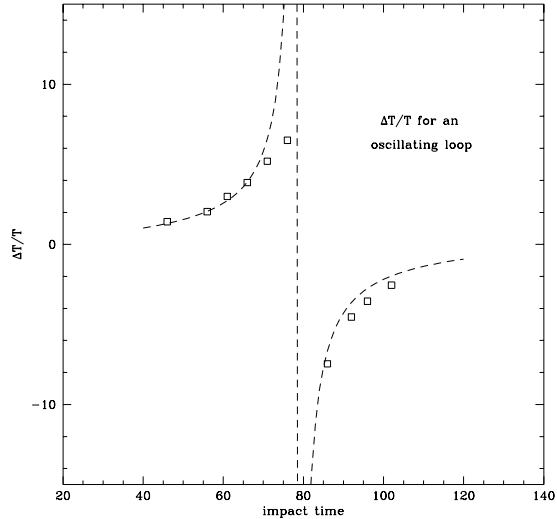

Figure 3: Temperature anisotropy from an oscillating cosmic string loop. The squares are the the temperature anisotropies measured from the simulation, and the dashed line is the analytic result.

In figure 4 we plot this standard deviation against gravity lattice box size. We find the variance of the maps is approximately constant from a box size of 20 up to 64. We can therefore be sure that with a box size of $64^3$, the results are not affected by the finite size of the lattice.

We ran the simulations with a gravity time step of $\Delta\eta = 1$. Halving this time step changed the final anisotropy maps locally by less than 1%, ensuring that we were in the quadratic convergence regime.

## 5 Results

Sample skies constructed using these methods for global monopoles, textures and nontopological textures can be found in reference [9], along with the power spectra for each theory. As expected, the power spectra show a striking suppression of power at high multipole moments, corresponding to the smoothing of small scale structure from reionisation.

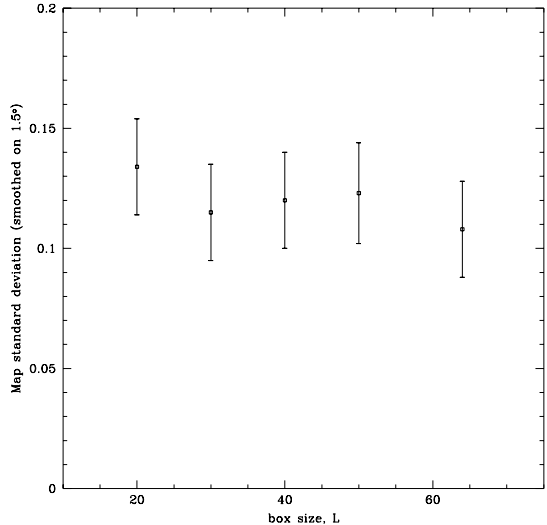

Figure 4: Standard deviation of a $(30°)^2$ temperature anisotropy map from a texture model as a function of lattice size. Error bars show the one sigma deviations for six maps.

The main focus of our work, however, was to determine whether the specific patterns expected from defect scenarios [15], [16] make them readily distinguishable from gaussian scenarios. Standard test of nongaussianity, such as skewness and kurtosis of the anisotropy patterns, proved to be statistically inconclusive.

Our method for determining whether a sky map is nongaussian or not is as follows. Given a map, we Fourier decompose the anisotropy. The phase of each mode is then randomised, and a map is created with the new Fourier components. This map has exactly the same power spectrum as the original map, although is now guaranteed to be gaussian (by the Central Limit Theorem). This procedure is repeated, constructing an ensemble of 1000 gaussian maps. We then searched for statistics which distinguish the original map from the gaussian ensemble.

Several authors have suggested that searching for large gradients might prove to be a useful way to distinguish nongaussian from gaussian theo-ries [17], [18]. We performed the following test. A map is considered "significantly nongaussian" if the maximum gradient in it is greater than that in 99% of the ensemble of Gaussian maps based upon it. Table 1 shows the number (out of eighteen) of maps satisfying this criterion for each theory (after smoothing with a Gaussian beam of FWHM $\theta_s$). Most of the texture and monopole maps were strongly nongaussian by this criterion, the non-topological texture much less so.

| Theory | $\theta_s = 0°$ | $\theta_s = 1.2°$ | $\theta_s = 2.4°$ | $\theta_s = 3.6°$ |
|---|---|---|---|---|
| Monopoles | 18 | 18 | 18 | 13 |
| Texture | 18 | 17 | 6 | 2 |
| NT Textures | 4 | 1 | 1 | 1 |

Table 1: The number of maps (out of 18) that were "significantly nongaussian", as defined in the text. (NT is non-topological.)

An obvious criticism of this test is its reliance on only a single point in the smoothed maps; one would prefer a statistic less sensitive to extreme data points. However, the significance of nongaussianity increases with increasing resolution, and it is the small scale structure of the relatively rare defect-induced signatures that is clearly the most significant feature. This suggests that the optimal experimental strategy to establish nongaussianity would be to measure temperature gradients at lower resolution (say, $1°$), and then focus in on regions of high gradient with higher resolution.

## 6 Conclusions

We have presented a method for calculating temperature anisotropy maps for global cosmic defect scenarios in a reionised universe. By explicitly considering the realistic photon paths, we have found a contribution to the Sachs-Wolfe formula which takes the form of a "Doppler" term, even in the case of baryons tracing the CDM exactly. We have also proposed a test which may distinguish the texture and monopole theories from gaussian theories, such as those based on inflation.

## 7 Acknowledgements

I would like to thank my collaborators from this project, Pedro Ferreira, Paul Graham , Ue-Li Pen and Neil Turok. I would also like to thank Lawrence

Krauss for organising this conference. The work presented here was partially funded by the UK EPSRC.